\documentclass[aps,showkeys,twocolumn,showpacs,preprintnumbers,amsmath,amssymb,superscriptaddress,floatfix,nofootinbib]{revtex4}
\usepackage{graphicx,color,dcolumn,booktabs,bm}
\usepackage{longtable,lscape}
\usepackage{txfonts}
\usepackage{overpic}
\usepackage{epsfig}
\usepackage{amssymb}
\usepackage{rotating}
\usepackage{epstopdf}
\usepackage{indentfirst}
\usepackage{feynmf}   %{feynmp}
\usepackage{slashed}  %for Feynman symbols
\usepackage{cases}
\usepackage{color}
\usepackage{multirow}
\usepackage{graphicx,color,dcolumn,booktabs,bm}
\usepackage{cases}
\usepackage{array}

\begin{document}

\title{The newly observed $\Lambda_b(6072)^0$ structure and its $\rho-$mode nonstrange partners }

\author{Wei Liang}
\affiliation{  Department
of Physics, Hunan Normal University,  Changsha 410081, China }

\affiliation{ Synergetic Innovation
Center for Quantum Effects and Applications (SICQEA), Changsha 410081,China}

\affiliation{  Key Laboratory of
Low-Dimensional Quantum Structures and Quantum Control of Ministry
of Education, Changsha 410081, China}

\author{Qi-Fang L\"u \footnote{Corresponding author} } \email{lvqifang@hunnu.edu.cn} %
\affiliation{  Department
of Physics, Hunan Normal University,  Changsha 410081, China }

\affiliation{ Synergetic Innovation
Center for Quantum Effects and Applications (SICQEA), Changsha 410081,China}

\affiliation{  Key Laboratory of
Low-Dimensional Quantum Structures and Quantum Control of Ministry
of Education, Changsha 410081, China}

\begin{abstract}
Inspired by the newly observed $\Lambda_b(6072)$ structure, we investigate its strong decay behaviors under various assignments within the $^3P_0$ model. Compared with the mass and total decay width, our results suggest that the $\Lambda_b(6072)$ can be regarded as the lowest $\rho-$mode excitation in $\Lambda_b$ family. Then, the strong decays of $\rho-$mode nonstrange partners for the $\Lambda_b(6072)$ are calculated. It is found that the $J^P=5/2^-$ $\Lambda_b$ and $\Lambda_c$ states are relatively narrow, and mainly decay into the $\Sigma_b^{(*)} \pi$ and $\Sigma_c^{(*)} \pi$ final states, respectively. These two states have good potentials to be observed in future experiments, which may help us to distinguish the three-quark model and diquark model.

\end{abstract}

\keywords{Strong decay; Bottom baryon; $\rho-$mode excitation}

\maketitle

\section{Introduction}
Recently, the CMS Collaboration observed an evidence for a broad excess of events in the $\Lambda_b^0 \pi^+ \pi^-$ mass spectrum in the region of $6040 \sim 6100$ MeV~\cite{Sirunyan:2020gtz}. Subsequently, the LHCb Collaboration also found a new baryon resonance $\Lambda_b(6072)^0$ in the same final states. The measured mass and decay width are~\cite{Aaij:2020rkw}:
\begin{eqnarray}
m[\Lambda_b(6072)^0] =6072.3\pm2.9~\pm0.6\pm0.2~\rm{MeV},
\end{eqnarray}
\begin{eqnarray}
\Gamma[\Lambda_b(6072)^0]= 72\pm11~\pm2~\rm{MeV},
\end{eqnarray}
which is consistent with the structure reported by the CMS Collaboration.

In recent years, a series of important progresses on the $\Lambda_{c(b)}$ and $\Sigma_{c(b)}$ families have been made experimentally. In 2017, a new charmed baryon $\Lambda_c(2860)$ was reported by the LHCb Collaboration in the $D^0 p$ final state, and the masses and decay widths of the $\Lambda_c(2880)$ and $\Lambda_c(2940)$ were measured~\cite{Aaij:2017vbw}. In 2018, the LHCb Collaboration found an excited bottom baryon $\Sigma_b(6097)$ in the $\Lambda^0_b \pi^\pm$ invariant mass spectrum~\cite{Aaij:2018tnn}. In 2019, the LHCb Collaboration reported the observation of $\Lambda_b(6146)$ and $\Lambda_b(6152)$ in the $\Lambda_b^0 \pi^+ \pi^-$ decay mode~\cite{Aaij:2019amv}, which were confirmed by the CMS Collaboration subsequently~\cite{Sirunyan:2020gtz}. These discoveries have attracted wide attentions of theorists, and plenty of works have been done to investigate the inner structures of observed $\Lambda_{c(b)}$ and $\Sigma_{c(b)}$ resonances~\cite{Chen:2017aqm,Wang:2017vtv,Yao:2018jmc,Lu:2018utx,Guo:2019ytq,Cui:2019dzj,Yang:2019cvw,Chen:2019ywy,Azizi:2020tgh,Chen:2018vuc,
Wang:2018fjm,Yang:2018lzg,Aliev:2018vye,Jia:2019bkr,Liang:2019aag,Wang:2019uaj,Chen:2016spr,Cheng:2015iom,Zhong:2007gp,Yoshida:2015tia,
Liang:2014eba,Lu:2016gev,He:2006is,Dong:2009tg,Luo:2019qkm,Cheng:2015naa,Ebert:2011kk,Roberts:2007ni,Chen:2014nyo,Wang:2017kfr,Cheng:2017ove,Garcilazo:2007eh,
Thakkar:2016dna,Ebert:2007nw,Capstick:1986bm}. In the traditional quark model, for the low-lying $\lambda-$mode $\Lambda_c$ states, the $\Lambda_c(2595)$ and $\Lambda_c(2625)$ resonances can be assigned as the $\Lambda_c(1P)$ doublet, the $\Lambda_c(2860)$ and $\Lambda_c(2880)$ resonances can be categorized into the $\Lambda_c(1D)$ doublet, and the $\Lambda_c(2765)$ can be regarded as the $\Lambda_c(2S)$ singlet ~\cite{Chen:2017aqm,Wang:2017vtv,Cheng:2017ove,Wang:2017kfr,Yao:2018jmc,Guo:2019ytq,Ebert:2011kk,Roberts:2007ni,Chen:2014nyo,Yoshida:2015tia}. Similarly, for the low-lying $\lambda-$mode $\Lambda_b$ states, the $\Lambda_b(5912)$ and $\Lambda_b(5920)$ have been assigned as the $\Lambda_b(1P)$ doublet, and the two structures $\Lambda_b(6146)$ and $\Lambda_b(6152)$ can be clarified into the $\Lambda_b(1D)$ doublet~\cite{Wang:2017kfr,Chen:2019ywy,Liang:2019aag,Wang:2019uaj,Ebert:2011kk,Roberts:2007ni,Chen:2014nyo,Ebert:2007nw,Capstick:1986bm}. It can be seen that the spectra of the low-lying $\lambda-$mode $\Lambda_c$ and $\Lambda_b$ states have been established well except for the $\Lambda_b(2S)$ state. For the $\Sigma_c$ and $\Sigma_b$ states, the similar patterns should also exist, and two $P-$wave candidates $\Sigma_c(2800)$ and $\Sigma_b(6097)$ have been observed experimentally~\cite{Mizuk:2004yu,Aaij:2018tnn}. Contrary to the abundant $\lambda-$mode states, no $\rho-$mode heavy baryon has been confirmed both theoretically and experimentally.

Considering the mass and width of $\Lambda_b(6072)$, the LHCb Collaboration suggested that it can be assigned as the first radial excitation of the $\Lambda_b$ baryon, the $\lambda-$mode $\Lambda_b(2S)$ state. Although the predicted mass of $\lambda-$mode $\Lambda_b(2S)$ state is consistent with the experimental observation~\cite{Ebert:2007nw,Capstick:1986bm,Roberts:2007ni,Ebert:2011kk,Chen:2014nyo}, the predicted total decay width is actually smaller than $72\pm11~\pm2~\rm{MeV}$ in the literature~\cite{Chen:2018vuc,Liang:2019aag,Wang:2019uaj}. Meanwhile, recent works suggested that it may be a Roper-like resonance~\cite{Arifi:2020yfp} or a overlap of two $\Sigma_b(1P)$ states~\cite{Xiao:2020gjo}. Before making a final conclusion, it is necessary to examine all possible interpretations carefully.

Based on the small predicted $\Lambda^0_b \pi \pi$ decay widths of $\Sigma_b(1P)$ states~\cite{Mu:2014iaa} and no signal with 6072 MeV in $\Lambda^0_b \pi^\pm$ decay modes~\cite{Aaij:2018tnn}, the LHCb Collaboration suggested that the interpretation of this newly observed structure as an excited $\Sigma_b(1P)$ resonance is disfavoured. However, for the $\Lambda^0_b \pi \pi$ decay mode, only the $\Sigma_b(1P) \to \Lambda^0_b (\pi \pi)_{I=1} \to \Lambda^0_b \pi \pi$ decay chain was calculated~\cite{Mu:2014iaa}, and other decay chains, such as $\Sigma_b(1P) \to \Sigma_b^{(*)} \pi \to \Lambda^0_b \pi \pi$, should also exist. Indeed, the decay chains via virtual heavy baryons play essential roles in the three-body decays of excited heavy baryons~\cite{Arifi:2020yfp,Arifi:2018yhr,Arifi:2017sac,Kawakami:2019hpp}. From Refs.~\cite{Liang:2019aag,Chen:2018vuc,Yang:2018lzg,Wang:2018fjm,Wang:2017kfr}, there are five $\Sigma_b(1P)$ states and several of them have rather small $\Lambda^0_b \pi^\pm$ branching ratios. It is possible for a $\Sigma_b(1P)$ state to be not observed in the $\Lambda^0_b \pi^\pm$ final states experimentally. Hence, the observed structure with 6072 MeV as $\Sigma_b(1P)$ assignments need further investigations.

In the three-quark model, the excitations of the heavy baryons can be divided into two parts, the $\rho-$mode and $\lambda-$mode. In the diquark model, the two light quarks are usually treated as a cluster without excitation, and only the $\lambda-$mode excitation exists. It can be seen that the diquark model freezes the degree of freedom between two light quarks and have less states than that of the three-quark model. In the relativized quark model, the predicted lowest $\rho-$mode $\Lambda_b$ state lies about 6100 MeV~\cite{Capstick:1986bm}, which is consistent with the experimental data. It is crucial to discuss the possibility of newly observed $\Lambda_b(6072)$ as a $\rho-$mode $\tilde{\Lambda}_b(1P)$ state. Moreover, it is a good opportunity to investigate these low-lying $\rho-$mode excitations in the nonstrange singly heavy baryon systems.

In this work, we tentatively assign the newly observed resonance $\Lambda_b(6072)$ as the $\lambda-$mode $\Lambda_b(2S)$, $\lambda-$mode $\Sigma_b(1P)$, and $\rho-$mode $\tilde{\Lambda}_b(1P)$ states, respectively, and calculate their strong decay behaviors. Our results indicate that the $\lambda-$mode $\Lambda_b(2S)$ and $\Sigma_b(1P)$ assignments are disfavored, while the $\Lambda_b(6072)$ as the lowest $\rho-$mode $\Lambda_b$ state is supported. With this interpretation, the $\Lambda_b(6072)$ may be the first observed $\rho-$mode excitation in singly heavy baryons. Then, the strong decays of the $\rho-$mode $\tilde{\Lambda}_c(1P)$, $\tilde{\Sigma}_b(1P)$ and $\tilde{\Sigma}_c(1P)$ states are calculated, and some of them have relatively narrow total decay widths. We hope these theoretical predictions on $\tilde{\Lambda}_{b(c)}(1P)$ and $\tilde{\Sigma}_{b(c)}(1P)$ states can provide valuable information for searching more $\rho-$mode excitations in future experiments.

The paper is organized as follows. In Sec.~\ref{Intoduction}, we introduce the $^3P_0$ model and notations briefly. In Sec.~\ref{interpretation}, the strong decays of the newly observed $\Lambda_b(6072)$ state under various assignments are investigated. The strong decay behaviors of the $\rho-$mode $\tilde{\Lambda}_c(1P)$, $\tilde{\Sigma}_b(1P)$ and $\tilde{\Sigma}_c(1P)$ states are presented in Sec.~\ref{prediction}. A summary is given in the last section.

\section{\label{Intoduction} Model and Notations}
In this issue, we adopt the $^3P_0$ model to study the strong decay behaviors of the singly heavy baryons. For a certain decay process, this model supposes that a quark-antiquark pair with the quantum number $J^{PC} =0^{++}$ is created from the vacuum and falls apart into the final states~\cite{Micu:1968mk}. The transition operator $T$ for a decay process $A \rightarrow BC$ can be expressed as
\begin{eqnarray}
T&=&-3\gamma\sum_m\langle 1m1-m|00\rangle\int
d^3\boldsymbol{p}_4d^3\boldsymbol{p}_5\delta^3(\boldsymbol{p}_4+\boldsymbol{p}_5)\nonumber\\
&&\times {\cal{Y}}^m_1\left(\frac{\boldsymbol{p}_4-\boldsymbol{p}_5}{2}\right
)\chi^{45}_{1,-m}\phi^{45}_0\omega^{45}_0b^\dagger_{4i}(\boldsymbol{p}_4)d^\dagger_{4j}(\boldsymbol{p}_5).
\end{eqnarray}
Here, the overall constant $\gamma$ reflects the $q_4\bar{q}_5$ pair-production strength, and
$\boldsymbol{p}_4$ and $\boldsymbol{p}_5$ are the momenta of the created quark-antiquark pair. The solid harmonic polynomial
${\cal{Y}}^m_1(\boldsymbol{p})\equiv|p|Y^m_1(\theta_p, \phi_p)$ is the $P-$wave distribution of the created quark pair in the momentum representation.  The $\phi^{45}_{0}$, $\omega^{45}_{0}$ and $\chi_{{1,-m}}^{45}$ are the flavor, color,
and spin parts of the quark pair $q_4\bar{q}_5$.

The transition matrix element for this model can be given by
\begin{eqnarray}
\langle
BC|T|A\rangle=\delta^3(\boldsymbol P_A - \boldsymbol P_B - \boldsymbol P_C){\cal{M}}^{M_{J_A}M_{J_B}M_{J_C}},
\end{eqnarray}
where the ${\cal{M}}^{M_{J_A}M_{J_B}M_{J_C}}$ corresponds to the helicity amplitude of the decay process $A\to B+C$. With the ${\cal{M}}^{M_{J_A}M_{J_B}M_{J_C}}$, one can calculate the partial width of this process directly
\begin{eqnarray}
\Gamma= \pi^2\frac{p}{M^2_A}\frac{1}{2J_A+1}\sum_{M_{J_A},M_{J_B},M_{J_C}}|{\cal{M}}^{M_{J_A}M_{J_B}M_{J_C}}|^2,
\end{eqnarray}
where $p=|\boldsymbol{p}|$ is the momentum of the final hadrons. More details of the $^3P_0$ model and the explicit formula can be found in Refs.~\cite{Chen:2007xf,Zhao:2016qmh,Chen:2016iyi,Lu:2019rtg}.

As mentioned above, there are two independent excitations in these baryon systems. The excitation between two light quarks is named as the $\rho-$mode, while the $\lambda-$mode stands for the excitation between the heavy quark and light quark subsystem. Then, the $P-$wave baryons can be divided into two groups, the $\rho-$mode states with $l_\rho=1$ and $l_\lambda=0$, and the $\lambda-$mode states with $l_\rho=0$ and $l_\lambda=1$. Here, we adopt a series of quantum numbers $n_\rho$, $l_\rho$, $n_\lambda$, $l_\lambda$, $L$, $S_\rho$, $j$, and $J^P$ to denote a theoretical state. The $n_\rho$ and $l_\rho$ are the radial and orbital quantum numbers between the two light quarks, respectively. Similarly, the $n_\lambda$ and $l_\lambda$ correspond to the radial and orbital quantum numbers between the heavy quark and light quark subsystem, respectively. The $S_\rho$ stands for the total spin of the two light quarks. The $j$ represents total angular momentum of the $L$ and $S_\rho$, where $L$ is the total orbital angular momentum. $J^P$ is the spin-parity of a heavy baryon.

The notations of initial states and predicted masses within the relativized quark model are listed in Table~\ref{mass}.  For the $\Lambda_b(6072)$ or $\Sigma_b(6072)$ resonance, we adopt its experimental mass under various assignments. For other $\rho-$mode baryons, we employ the theoretical masses to calculate their strong decays. All the parameters adopted here are the same as our previous works, which have been widely used for the strong decays of singly heavy baryon systems~\cite{Liang:2019aag,Liang:2020hbo,Lu:2019rtg,Lu:2018utx,Lu:2020ivo}.

\begin{table*}[htb]
\begin{center}
\caption{ \label{mass} Notations, quantum numbers and masses of the initial baryons. The masses are taken from the relativized quark model~\cite{Capstick:1986bm}. The $Q$ stands for a charm or bottom quark, and the $\sim$ is used for the $\rho-$mode states. The units are in MeV. }
\renewcommand{\arraystretch}{1.5}
\begin{tabular*}{18cm}{@{\extracolsep{\fill}}*{12}{p{1.4cm}<{\centering}}}
\hline\hline
State & $n_{\rho}$ & $l_{\rho}$ & $n_{\lambda}$ & $l_{\lambda}$ & $L$ & $S_{\rho}$ & $j$ & $J^{P}$ &Charmed &Bottom \tabularnewline
\hline
$\text{\ensuremath{\Lambda_{Q}(2S)}}$ & 0 & 0 & 1 & 0 & 0 & 0 & 0 & $\frac{1}{2}^{+}$ & 2775 & 6045\tabularnewline
$\text{\ensuremath{\tilde{\Lambda}_{Q0}(\frac{1}{2}^{-})}}$ & 0 & 1 & 0 & 0 & 1 & 1 & 0 & $\frac{1}{2}^{-}$ & 2780 & 6100\tabularnewline
$\text{\ensuremath{\tilde{\Lambda}_{Q1}(\frac{1}{2}^{-})}}$ & 0 & 1 & 0 & 0 & 1 & 1 & 1 & $\frac{1}{2}^{-}$ & 2830 & 6165\tabularnewline
$\text{\ensuremath{\tilde{\Lambda}_{Q1}(\frac{3}{2}^{-})}}$ & 0 & 1 & 0 & 0 & 1 & 1 & 1 & $\frac{3}{2}^{-}$ & 2840 & 6185\tabularnewline
$\text{\ensuremath{\tilde{\Lambda}_{Q2}(\frac{3}{2}^{-})}}$ & 0 & 1 & 0 & 0 & 1 & 1 & 2 & $\frac{3}{2}^{-}$ & 2885 & 6190\tabularnewline
$\text{\ensuremath{\tilde{\Lambda}_{Q2}(\frac{5}{2}^{-})}}$ & 0 & 1 & 0 & 0 & 1 & 1 & 2 & $\frac{5}{2}^{-}$ & 2900 & 6205\tabularnewline
$\text{\ensuremath{\Sigma_{Q0}(\frac{1}{2}^{-})}}$ & 0 & 0 & 0 & 1 & 1 & 1 & 0 & $\frac{1}{2}^{-}$ & 2765 & 6070\tabularnewline
$\text{\ensuremath{\text{\ensuremath{\Sigma_{Q1}(\frac{1}{2}^{-})}}}}$ & 0 & 0 & 0 & 1 & 1 & 1 & 1 & $\frac{1}{2}^{-}$ & 2770 & 6070\tabularnewline
$\text{\ensuremath{\Sigma_{Q1}(\frac{3}{2}^{-})}}$ & 0 & 0 & 0 & 1 & 1 & 1 & 1 & $\frac{3}{2}^{-}$ & 2770 & 6070\tabularnewline
$\text{\ensuremath{\Sigma_{Q2}(\frac{3}{2}^{-})}}$ & 0 & 0 & 0 & 1 & 1 & 1 & 2 & $\frac{3}{2}^{-}$ & 2805 & 6085\tabularnewline
$\text{\ensuremath{\Sigma_{Q2}(\frac{5}{2}^{-})}}$ & 0 & 0 & 0 & 1 & 1 & 1 & 2 & $\frac{5}{2}^{-}$ & 2815 & 6090\tabularnewline
$\text{\ensuremath{\tilde{\Sigma}_{Q1}(\frac{1}{2}^{-})}}$ & 0 & 1 & 0 & 0 & 1 & 0 & 1 & $\frac{1}{2}^{-}$ & 2840 & 6170\tabularnewline
$\text{\ensuremath{\text{\ensuremath{\tilde{\Sigma}_{Q1}(\frac{3}{2}^{-})}}}}$ & 0 & 1 & 0 & 0 & 1 & 0 & 1 & $\frac{3}{2}^{-}$ & 2865 & 6180\tabularnewline
\hline\hline
\end{tabular*}
\end{center}
\end{table*}

\section{\label{interpretation} Strong decays of the $\Lambda_b(6072)$ or $\Sigma_b(6072)$ }
\subsection{$\Lambda_b(2S)$}
As mentioned in the introduction, LHCb Collaboration suggested that the $\Lambda_b(6072)$ may be regarded as $\Lambda_b(2S)$ state. The predicted masses of the $\Lambda_b(2S)$ state within quark models are also consistent with the experimental data. The strong decays of this assignment are calculated and listed in Table~\ref{Lambda_b 2S}. The predicted width is about 9 MeV, which is significantly smaller than the experimental observation. Thus, our calculations do not support the $\Lambda_b(6072)$ as $\Lambda_b(2S)$ state.

\begin{table}[htb]
\begin{center}
\caption{ \label{Lambda_b 2S} Strong decays of the $\Lambda_b(6072)$ as $\Lambda_b(2S)$ state in MeV.}
\renewcommand{\arraystretch}{1.5}
\begin{tabular*}{8.5cm}{@{\extracolsep{\fill}}*{2}{p{3.5cm}<{\centering}}}
\hline\hline
Decay mode & $\text{\ensuremath{\Lambda_{Q}(2S)}}$\tabularnewline
\hline
$\Sigma_{b}\pi$ & 3.93\tabularnewline
$\Sigma_{b}^{*}\pi$ & 5.34\tabularnewline
Total width & 9.27\tabularnewline
Experiment & $72\pm11\pm2$\tabularnewline
\hline\hline
\end{tabular*}
\end{center}
\end{table}

Other theoretical works also predicted smaller total decay widths of the $\Lambda_b(2S)$ state, which varies in the range of $2 \sim 36$ MeV~\cite{Chen:2018vuc,Liang:2019aag,Wang:2019uaj}. These calculations indicate that the $\Lambda_b(2S)$ should be a narrow state. Both $\Sigma_b \pi$ and $\Sigma_b^* \pi$ decay modes for the $\Lambda_b(2S)$ state are important, which can be tested by future experiments.

\subsection{$\Sigma_b(1P)$}
In the traditional quark model, there are five $\lambda-$mode $\Sigma_b(1P)$ states, and their masses are predicted to be around  6070 $\thicksim$ 6090 MeV in relativized quark model. It is possible to regard the newly observed structure as the $\Sigma_b(1P)$ states. The strong decay behaviors of $\Sigma_b(1P)$ assignments are presented in Table~\ref{Sigma_b 1P}. It is shown that all the pure $\Sigma_b(1P)$ assignments are disfavored.

\begin{table*}[htb]
\begin{center}
\caption{ \label{Sigma_b 1P} Decay widths of the $\Sigma_b(6072)$ as $\Sigma_b(1P)$ states in MeV.}
\renewcommand{\arraystretch}{1.5}
\begin{tabular*}{18cm}{@{\extracolsep{\fill}}*{11}{p{1.8cm}<{\centering}}}
\hline\hline
Decay mode & $\text{\ensuremath{\Sigma_{b0}(\frac{1}{2}^{-})}}$ & $\text{\ensuremath{\Sigma_{b1}(\frac{1}{2}^{-})}}$ & $\text{\ensuremath{\Sigma_{b1}(\frac{3}{2}^{-})}}$ & $\text{\ensuremath{\Sigma_{b2}(\frac{3}{2}^{-})}}$ & $\text{\ensuremath{\Sigma_{b2}(\frac{5}{2}^{-})}}$\tabularnewline
\hline
$\Lambda_{b}\pi$ & 263.54 & $\cdots$ & $\cdots$ & 8.65 & 8.65\tabularnewline
$\Sigma_{b}\pi$ & $\cdots$ & 208.93 & 0.22 & 0.40 & 0.18\tabularnewline
$\Sigma_{b}^{*}\pi$ & $\cdots$ & 0.23 & 180.35 & 0.21 & 0.33\tabularnewline
Total width & 263.54 & 209.16 & 180.57 & 9.26 & 9.16\tabularnewline
Experiment & \multicolumn{5}{c}{$72\pm11\pm2$}\tabularnewline
\hline\hline
\end{tabular*}
\end{center}
\end{table*}

\begin{figure}[!htbp]
\includegraphics[scale=0.21]{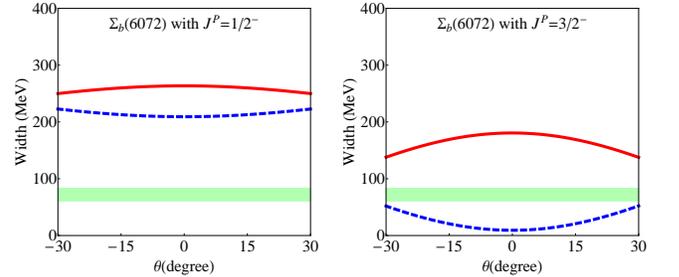}
\vspace{0.0cm} \caption{\label{mixing1}
The total decay widths versus the mixing angle $\theta$ . The red solid and blue dashed lines correspond to the $|1P~{J^P}\rangle_1$ and $|1P~{J^P}\rangle_2$ states, respectively, where the $J^P$ equals to $1/2^-$ or $3/2^-$. The green bands are the experimental total decay widths.}
\end{figure}

However, the physical resonance may correspond to the mixing of the quark model states. The mixing scheme of $P-$wave states can be expressed as
\[\binom{\left\vert 1P~1/2^{-}\right\rangle _{1}}{\left\vert
1P~1/2^{-}\right\rangle _{2}}=\left(\begin{array}{cc} \cos\theta  & \sin \theta  \\ -\sin \theta  & \cos \theta
\end{array}%
\right) \left(\begin{array}{c} \left\vert 1/2^{-},j=0\right\rangle  \\
\left\vert 1/2^{-},j=1\right\rangle
\end{array}%
\right) ,\]%
\[
\binom{\left\vert 1P~3/2^{-}\right\rangle _{1}}{\left\vert
1P~3/2^{-}\right\rangle _{2}}=\left(\begin{array}{cc}
\cos \theta  & \sin \theta  \\ -\sin \theta & \cos \theta
\end{array}%
\right) \left( \begin{array}{c}
\left\vert 3/2^{-},j=1\right\rangle  \\
\left\vert 3/2^{-},j=2\right\rangle
\end{array}%
\right) .
\]
Due to the finite mass of bottom quark, a small mixing angle is allowed which breaks the heavy quark symmetry explicitly. The total decay widths of various assignments versus the mixing angle $\theta$ in the range of $-30^\circ \sim 30^\circ$ are plotted in Figure~\ref{mixing1}. It can be seen that when the mixing angle $\theta$ goes close to $\pm 30^\circ$, the predicted width of $\Sigma_b(6072)$ is consistent with the experimental data. This mixing angle seems larger than that of previous works on the singly bottom baryons~\cite{Liang:2020hbo,Liang:2019aag}. With this mixing angle, the calculated total width of $\Sigma_b(6072)$ is about 52 MeV and the predicted branching ratios of dominating channels are
\begin{equation}
Br(\Lambda_b \pi, \Sigma^*_b \pi)=12.4\%,  86.4\%,
\end{equation}
which indicates that the branching ratio of $\Lambda_b \pi$ final state is also significant. However, the LHCb Collaboration did not observe the $\Sigma_b(6072)$ signal in $\Lambda_b \pi$ mass spectrum~\cite{Aaij:2018tnn}. Moreover, there already exists a good candidate $\Sigma_b(6097)$ for the $|1P~{3/2^-}\rangle_2$ state in the literatures~\cite{Aaij:2018tnn,Liang:2019aag,Cui:2019dzj,Chen:2018vuc,Wang:2018fjm,Yang:2018lzg,Aliev:2018vye,Jia:2019bkr}. Given the large mixing angle and significant branching ratio of $\Lambda_b \pi$ channel, the newly observed structure as $\Sigma_b(1P)$ assignments are disfavored.

\subsection{$\tilde{\Lambda}_b(1P)$}
According to the quark model, there are five $\rho-$mode $\tilde{\Lambda}_b(1P)$ states, and the predicted masses within relativized quark model vary from 6100 to 6205 MeV. Considering the uncertainties of quark model, it is possible to treat the $\Lambda_b(6072)$ as the lowest one of $\tilde{\Lambda}_b(1P)$ states. Meanwhile, the masses for the $\rho-$mode $\tilde{\Lambda}_b(1P)$ states can be estimated with the help of the $\lambda-$mode resonances. In the three-quark model with harmonic oscillator approximation~\cite{Zhong:2007gp}, the masses between $\rho-$mode and $\lambda-$mode $P-$wave excitations can be expressed as
\begin{eqnarray}
\frac{\bar m[\tilde{\Lambda}_b(1P)] - m[\Lambda_b(1S)]}{\bar m[\Lambda_b(1P)] - m[\Lambda_b(1S)]} = \sqrt{\frac{3m_b}{2m_q+m_b}},
\end{eqnarray}
where the $\bar m[\tilde{\Lambda}_b(1P)]$ and $\bar m[\Lambda_b(1P)]$ are the average masses of $\tilde{\Lambda}_b(1P)$ and $\Lambda_b(1P)$ states, respectively. With the masses of $\Lambda_b(5912)$ and $\Lambda_b(5920)$, one can obtain
\begin{small}
\begin{eqnarray}
\small\bar m[\Lambda_b(1P)] = \frac{2m[\Lambda_b(5912)]+4m[\Lambda_b(5920)]}{6}=5917.35~\rm{MeV}.
\end{eqnarray}
\end{small}
The $m_b=4977~\rm{MeV}$ and $m_q=220~\rm{MeV}$ are the masses of bottom and light quarks, respectively~\cite{Liang:2019aag,Liang:2020hbo}. Then, the estimated $\bar m[\tilde{\Lambda}_b(1P)]$ equals to 6114 MeV, which is consistent with the predicted average mass from relativized quark model~\cite{Capstick:1986bm}. When the mass splitting is considered, the lowest $\tilde{\Lambda}_b(1P)$ state should lie around 6072 MeV. Moreover, this estimation is not sensitive to the parameters of quark masses because the bottom quark is much heavier than the light quark and the ratio approximately equals to $\sqrt{3}$.

Here, the strong decays of the $\tilde{\Lambda}_b(1P)$ assignments are calculated and shown in Table~\ref{Lambda_b 1P1}. For the pure $j=0$ state, the strong decays are forbidden due to the quantum number conservation. The predicted widths of two $j=1$ states are extremely large, while the two $j=2$ assignments show smaller total decay widths. Our results indicate that the $\Lambda_b(6072)$ as pure  $\tilde{\Lambda}_b(1P)$ states can be excluded.
\begin{table*}[htb]
\begin{center}
\caption{ \label{Lambda_b 1P1} Strong decays of the $\Lambda_b(6072)$ as $\tilde{\Lambda}_b(1P)$ states in MeV.}
\renewcommand{\arraystretch}{1.5}
\begin{tabular*}{18cm}{@{\extracolsep{\fill}}*{11}{p{1.8cm}<{\centering}}}
\hline\hline
Decay mode & $\text{\ensuremath{\tilde{\Lambda}_{b0}(\frac{1}{2}^{-})}}$ & $\text{\ensuremath{\tilde{\Lambda}_{b1}(\frac{1}{2}^{-})}}$ & $\text{\ensuremath{\tilde{\Lambda}_{b1}(\frac{3}{2}^{-})}}$ & $\text{\ensuremath{\tilde{\Lambda}_{b2}(\frac{3}{2}^{-})}}$ & $\text{\ensuremath{\tilde{\Lambda}_{b2}(\frac{5}{2}^{-})}}$\tabularnewline
\hline
$\Sigma_{b} \pi$ & $\cdots$ & 521.81 & 1.38 & 2.47 & 1.10\tabularnewline
$\Sigma_{b}^{*}\pi$ & $\cdots$ & 1.47 & 458.33 & 1.30 & 2.03\tabularnewline
Total width & $\cdots$ & 523.28 & 459.71 & 3.77 & 3.13\tabularnewline
Experiment & \multicolumn{5}{c}{$72\pm11\pm2$}\tabularnewline
\hline\hline
\end{tabular*}
\end{center}
\end{table*}

\begin{figure}[!htbp]
\includegraphics[scale=0.21]{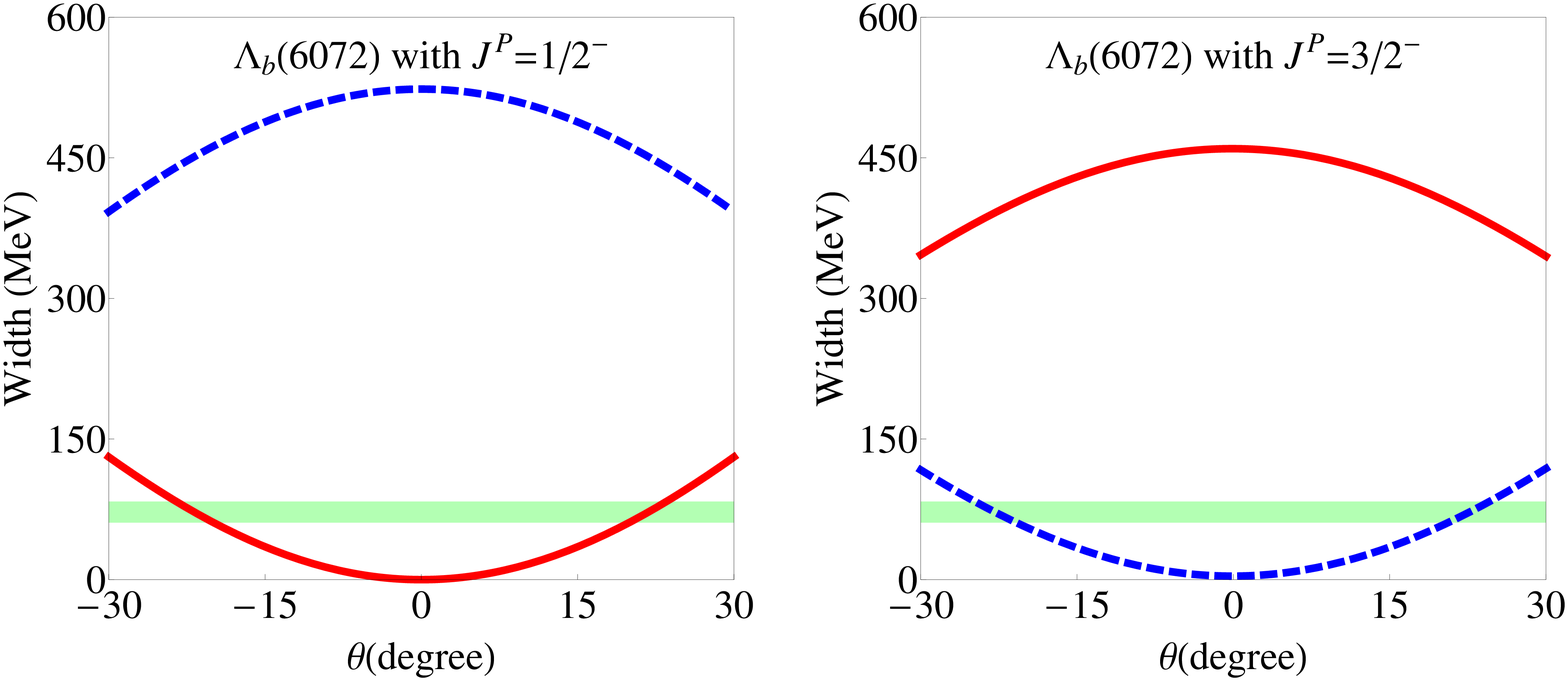}
\vspace{0.0cm} \caption{\label{mixing2} The total decay widths versus the mixing angle $\theta$ . The red solid and blue dashed lines correspond to the $|1P~{J^P}\rangle_1$ and $|1P~{J^P}\rangle_2$ states, respectively, where the $J^P$ equals to $1/2^-$ or $3/2^-$. The green bands are the experimental total decay widths.}
\end{figure}

The mixing of the states with same $J^P$ is taken into account. The mixing scheme is similar to the $\Sigma_b(1P)$ states, and the results are shown in Figure~\ref{mixing2}. Obviously, the $\Lambda_b(6072)$ state can be regarded as the $J^P=1/2^-$ $\tilde \Lambda_b(1P)$ state with large $j=0$ component. Meanwhile, the strong decays also support it as the $J^P=3/2^-$ $\tilde \Lambda_b(1P)$ state with large $j=2$ component. From Table~\ref{mass}, the predicted masses of $\tilde{\Lambda}_{b0}(\frac{1}{2}^{-})$ and $\tilde{\Lambda}_{b2}(\frac{3}{2}^{-})$ states are 6100 and 6190 MeV, respectively. Given these predicted masses, the $|1P~1/2^- \rangle_1$ assignment is favored, while the $|1P~3/2^- \rangle_2$ assignment can be excluded.

\begin{figure}[!htbp]
\includegraphics[scale=0.21]{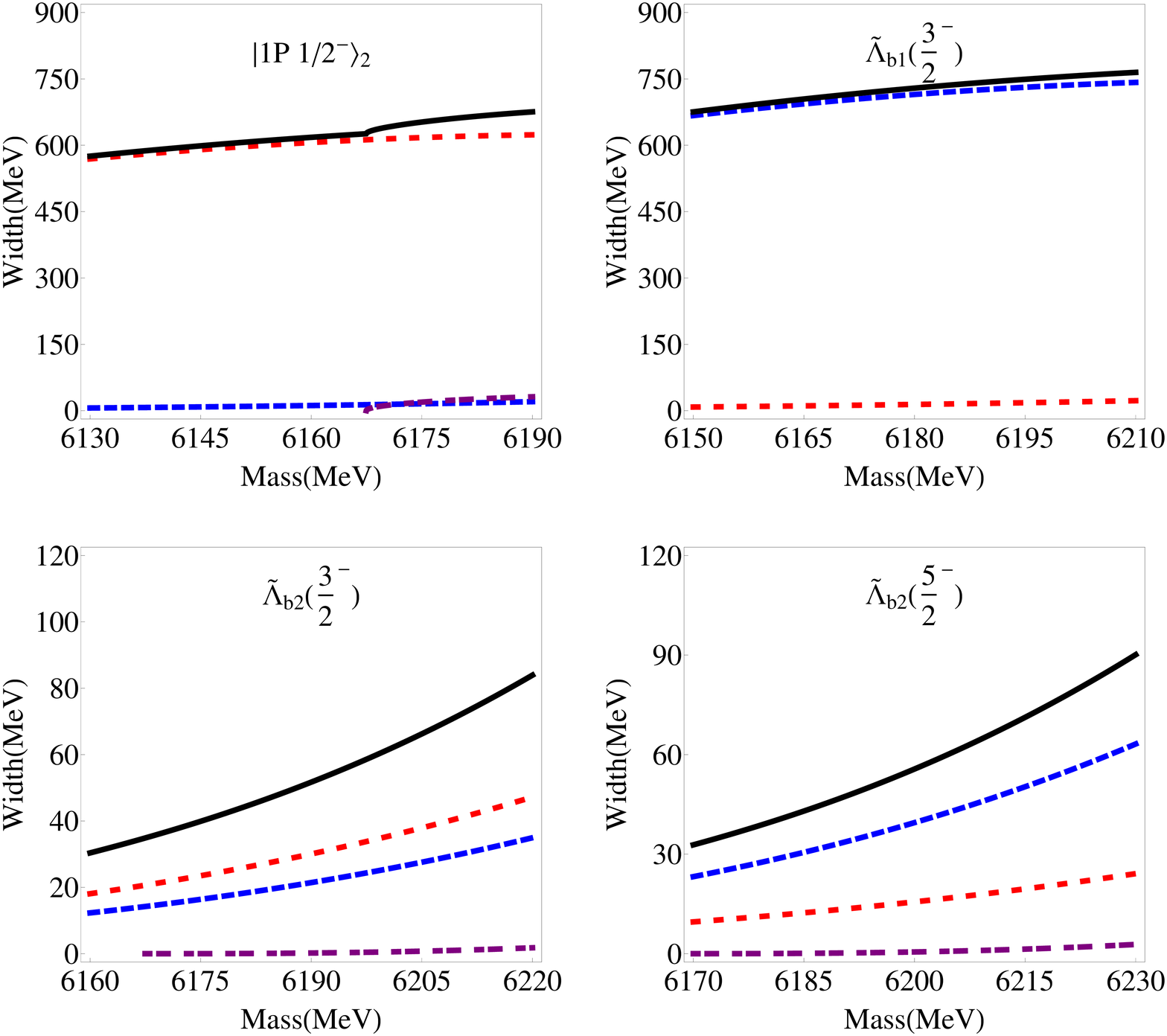}
\vspace{0.0cm} \caption{\label{mass1} The decay behaviors for the partners of $\Lambda(6072)$ versus their initial masses. The black lines correspond to the total decay widths. The red dotted, blue dashed, purple dot-dashed lines stand for the partial decay widths of $\Sigma_{b} \pi$, $\Sigma_{b}^{*} \pi$, and $\Lambda_{b} \eta$ channels, respectively. The $\Lambda_b \eta$ mode for $|1P~1/2^- \rangle_2$ state arises due to the allowed phase space and mixing scheme.}
\end{figure}

When the mixing angle is $22^\circ$, the strong decay width of $|1P~1/2^- \rangle_1$ state is about 73 MeV, which is consistent with the experimental data. The branching ratios are predicted to be
\begin{equation}
Br(\Sigma_b \pi, \Sigma^*_b \pi)=99.7\%, 0.3\%,
\end{equation}
which indicates that the $\Sigma_b \pi$ is the predominant decay channel and the three-body decay can occur via the $\Lambda(6072) \to \Sigma_b \pi \to \Lambda_b \pi \pi$ process. For the other states, we recalculate their strong decays around their predicated masses and present the results in Figure~\ref{mass1}. For the $|1P~1/2^- \rangle_2$ and $\tilde{\Lambda}_{b1}(\frac{3}{2}^{-})$ states, it is difficult to find them experimentally due to the large total decay widths. However, for the $\tilde{\Lambda}_{b2}(\frac{3}{2}^{-})$ and $\tilde{\Lambda}_{b2}(\frac{5}{2}^-)$ states, their total widths are relatively narrow, which have good potentials to be observed by future experiments. Also, when the initial masses for the $j=2$ states increase, the total decay widths become larger and the dominating channels remain.

It should be mentioned that the $\tilde{\Lambda}_{b2}(\frac{5}{2}^-)$ state is a crucial point for distinguishing the three-quark model and diquark model. In the three-quark model, the lowest $J^P=5/2^-$ $\Lambda_b$ state belongs to the $\rho-$mode $P-$wave excitation. However, the $\rho-$mode excitations are frozen in the diquark model, and the lowest $J^P=5/2^-$ state is the $F-$wave $\Lambda_{b2}(\frac{5}{2}^-)$. The predicated masses of $\Lambda_{b2}(\frac{5}{2}^-)$ in diquark model lie in the region of $6346 \sim 6408$ MeV~\cite{Ebert:2007nw,Ebert:2011kk,Chen:2014nyo}, which is significantly larger than that of $\tilde \Lambda_{b2}(\frac{5}{2}^-)$ state in the three-quark model~\cite{Capstick:1986bm,Roberts:2007ni}. The future experimental searches can help us to distinguish these two types of models.

\section{\label{prediction} Strong decays of nonstrange partners }
\subsection{$\tilde{\Lambda}_c(1P)$}
\begin{figure}[!htbp]
\includegraphics[scale=0.21]{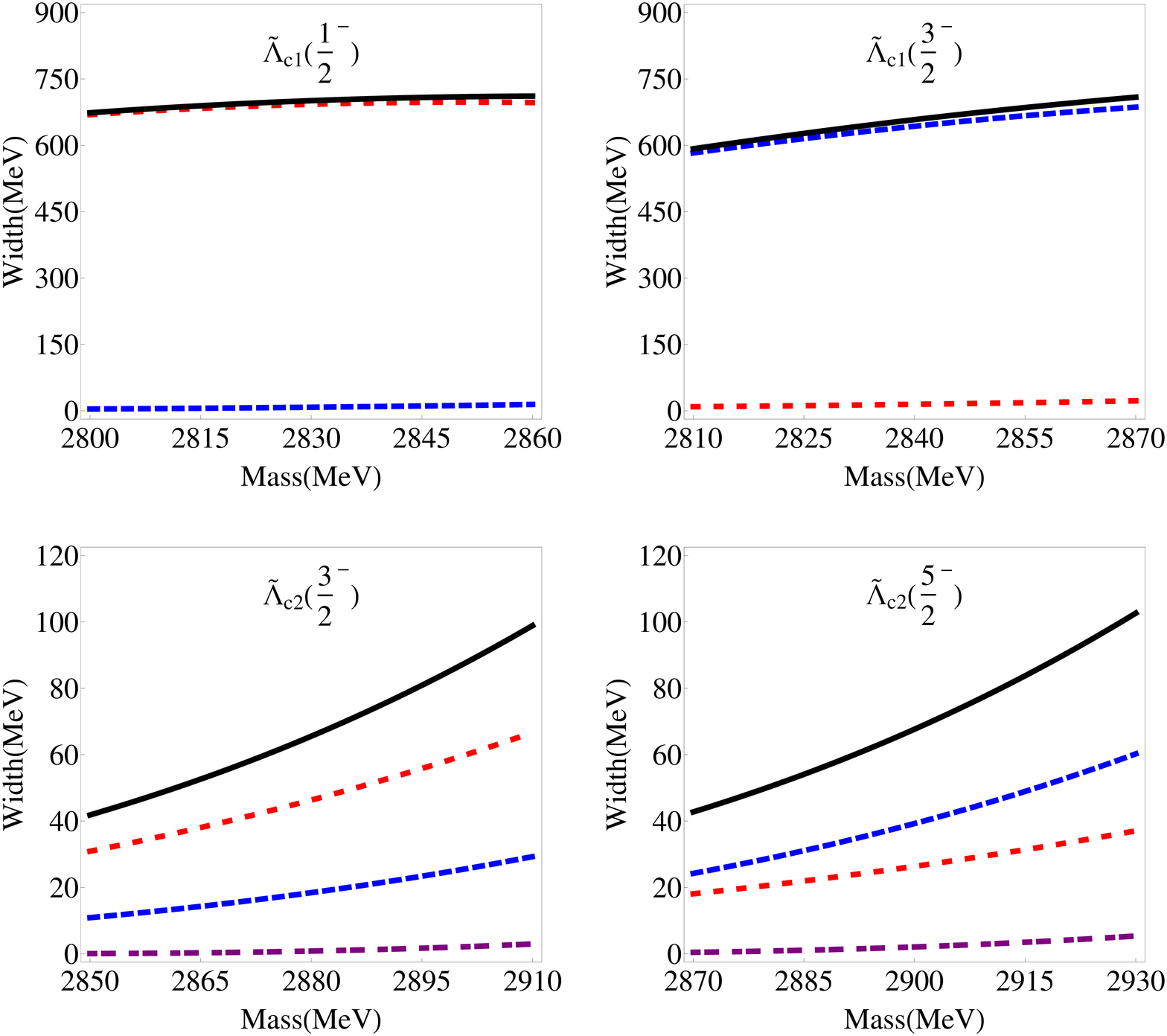}
\vspace{0.0cm} \caption{\label{mass2} The decay behaviors for the pure $j=1$ and $j=2$ $\tilde{\Lambda}_c(1P)$ states versus their initial masses. The black lines correspond to the total decay widths. The red dotted, blue dashed, purple dot-dashed lines stand for the partial decay widths of $\Sigma_{c} \pi$, $\Sigma_{c}^{*} \pi$, and $\Lambda_{c} \eta$ channels, respectively.}
\end{figure}

In the constituent quark model, there are five $\rho-$mode $\tilde{\Lambda}_c(1P)$ states, and their predicted masses in relativized quark model are about $2780 \sim 2900$ MeV. Similarly, one can estimate the average mass of $\rho-$mode $\tilde{\Lambda}_c(1P)$ states with the help of the $\lambda-$mode excitations. The average mass $\bar m[\tilde{\Lambda}_c(1P)]$ equals to 2793 MeV, which is consistent with the result of relativized quark model~\cite{Capstick:1986bm}.   Within these masses, their strong decays are calculated and listed in Table~\ref{Lambda_c 1P1}. For the $j=0$ state, the OZI-allowed strong decays are forbidden due to the limited phase space and quantum number conservation. For the two $j=1$ states, their total decay width are predicted to be several hundred MeV, and the main decay modes for $\tilde{\Lambda}_{c1}(\frac{1}{2}^-)$ and $\tilde{\Lambda}_{c1}(\frac{3}{2}^-)$ states are $\Sigma_c \pi$ and $\Sigma_c^* \pi$, respectively. Due to the large decay widths, these two states can hardly be observed experimentally. For the two $j=2$ states, the total decay widths for the $J^P=3/2^-$ and $5/2^-$ states are about 70 and 68 MeV, respectively. The main decay channels are $\Sigma_c \pi$ and $\Sigma_c^* \pi$, and the branching ratios are
\begin{equation}
Br(\Sigma_c \pi, \Sigma^*_c \pi)=70.1\%, 28.4\%, \qquad J^P=3/2^-,
\end{equation}
\begin{equation}
Br(\Sigma_c \pi, \Sigma^*_c \pi)=38.9\%, 58.0\%, \qquad J^P=5/2^-.
\end{equation}
In Ref.~\cite{Chen:2014nyo}, the authors claimed that the lowest $J^P=5/2^-$ $\Lambda_c$ state is a nice criterion to test the three-quark model and diquark model for charmed baryons. Our results indicate that the $\tilde{\Lambda}_{c2}(\frac{5}{2}^{-})$ state is not a broad state and can be searched in the $\Sigma_c \pi$ and $\Sigma_c^* \pi$ final states. Also, we plot the decay widths as functions of the initial masses for the pure $j=1$ and 2 states in Figure~\ref{mass2}. It can be seen that the total decay widths increases with the rise of initial masses, but the branching ratios are almost unchanged.

\begin{table*}[htb]
\begin{center}
\caption{ \label{Lambda_c 1P1}Strong decays of the $\tilde{\Lambda}_c(1P)$ states with the predicted masses in MeV. }
\renewcommand{\arraystretch}{1.5}
\begin{tabular*}{18cm}{@{\extracolsep{\fill}}*{11}{p{1.8cm}<{\centering}}}
\hline\hline
Decay mode & $\text{\ensuremath{\tilde{\Lambda}_{c0}(\frac{1}{2}^{-})}}$ & $\text{\ensuremath{\tilde{\Lambda}_{c1}(\frac{1}{2}^{-})}}$ & $\text{\ensuremath{\tilde{\Lambda}_{c1}(\frac{3}{2}^{-})}}$ & $\text{\ensuremath{\tilde{\Lambda}_{c2}(\frac{3}{2}^{-})}}$ & $\text{\ensuremath{\tilde{\Lambda}_{c2}(\frac{5}{2}^{-})}}$\tabularnewline
\hline
$\Sigma_{c} \pi$ & $\cdots$ & 692.79 & 14.84 & 49.35 & 26.35\tabularnewline
$\Sigma_{c}^{*} \pi$ & $\cdots$ & 8.15 & 643.21 & 19.97 & 39.25\tabularnewline
$\Lambda_{c} \eta$ & $\cdots$ & $\cdots$ & $\cdots$ & 1.07 & 2.06\tabularnewline
Total width & $\cdots$ & 700.94 & 658.05 & 70.39 & 67.66\tabularnewline
\hline\hline
\end{tabular*}
\end{center}
\end{table*}

By assuming various observed $\Lambda_c$ baryons as the $\tilde{\Lambda}_{c}$ states, the authors discussed their strong decay behaviors~\cite{Guo:2019ytq}, where the calculated branching ratios are consistent with ours. It can be also seen that the strong decays for the $\tilde{\Lambda}_b(1P)$ and $\tilde{\Lambda}_c(1P)$ states show similar features, which preserve the heavy quark symmetry approximately. Certainly, the physical resonances can be mixing of the theoretical states in the quark model, and the mixing scheme is similar with the $\tilde{\Lambda}_b(1P)$ case. With a small mixing angle, the $|1P~1/2^- \rangle_1$ state in the charm sector should be relatively narrow, which can be searched in the $\Sigma_c \pi$ final state in future.

\subsection{$\tilde{\Sigma}_b(1P)$}
From Table~\ref{mass}, two $\rho-$mode $\tilde{\Sigma}_b(1P)$ states exist in the conventional quark model, and their masses are predicted to be 6170 MeV and 6180 MeV in the relativized quark model. With the calculated masses, their strong decays are estimated and listed in Table~\ref{Sigma_b 1P1}. The total decay widths for these two states are about 260 MeV, which are rather large. The main decay modes are $\Sigma_b \pi$ and $\Sigma_b^* \pi$ for the $J^P=1/2^-$ and $3/2^-$ states, respectively. The branching ratios for these two states are
\begin{equation}
Br(\Sigma_b \pi, \Sigma^*_b \pi)=91.3\%, 8.7\%, \qquad J^P=1/2^-,
\end{equation}
\begin{equation}
Br(\Sigma_b \pi, \Sigma^*_b \pi)=7.2\%, 92.8\%, \qquad J^P=3/2^-.
\end{equation}
In Ref.~\cite{Yang:2018lzg}, by assuming the $\Sigma_b(6097)$ as the $\tilde{\Sigma}_b(1P)$ states, the authors estimated the strong decays, and the calculated total decay widths are about 168-235 MeV. Since large masses are adopted for the initial $\tilde \Sigma_b(1P)$ states in present work, the predicted decay widths are also larger than the previous work~\cite{Yang:2018lzg}. More theoretical efforts on masses and decays are needed to understand these two $\tilde{\Sigma}_b(1P)$ states.

\begin{table}[htb]
\begin{center}
\caption{ \label{Sigma_b 1P1}Strong decays of the $\tilde{\Sigma}_b(1P)$ states with the predicted masses in MeV.}
\renewcommand{\arraystretch}{1.5}
\begin{tabular*}{8.5cm}{@{\extracolsep{\fill}}*{3}{p{2.5cm}<{\centering}}}
\hline\hline
Decay mode & $\text{\ensuremath{\tilde{\Sigma}_{b1}(\frac{1}{2}^{-})}}$ & $\text{\ensuremath{\tilde{\Sigma}_{b1}(\frac{3}{2}^{-})}}$\tabularnewline
\hline
$\Sigma_{b} \pi$ & 238.43 & 19.44\tabularnewline
$\Sigma_{b}^{*} \pi$ & 22.64 & 248.84\tabularnewline
Total width & 261.07 & 268.28\tabularnewline
\hline\hline
\end{tabular*}
\end{center}
\end{table}

\subsection{$\tilde{\Sigma}_c(1P)$}
The predicted masses of the $\tilde{\Sigma}_{c1}(\frac{1}{2}^-)$ and $\tilde{\Sigma}_{c1}(\frac{3}{2}^-)$ states in the relativized quark model are 2840 and 2865 MeV, respectively. Their calculated decay widths are presented in Table~\ref{Sigma_c 1P1}. It is shown that the total decay widths are quite large and the main decay modes are $\Sigma_c \pi$ and $\Sigma^*_c \pi$ for the $J^P=1/2^-$ and $3/2^-$ states, respectively. These results are similar with the $\tilde{\Sigma}_b(1P)$ states, which preserves the heavy quark symmetry well. In Ref.~\cite{Chen:2007xf}, the authors calculated the strong decays of the $\tilde{\Sigma}_c(1P)$ states with a mass of 2882 MeV, and the calculated total decay width are also large. However, the predicted branching ratios in present work are quite different with theirs, and more theoretical works are demanded to clarify this problem.

\begin{table}[htb]
\begin{center}
\caption{ \label{Sigma_c 1P1}Strong decays of the $\tilde{\Sigma}_c(1P)$ states with the predicted masses in MeV. }
\renewcommand{\arraystretch}{1.5}
\begin{tabular*}{8.5cm}{@{\extracolsep{\fill}}*{3}{p{2.5cm}<{\centering}}}
\hline\hline
Decay mode & $\text{\ensuremath{\tilde{\Sigma}_{c1}(\frac{1}{2}^{-})}}$ & $\text{\ensuremath{\tilde{\Sigma}_{c1}(\frac{3}{2}^{-})}}$\tabularnewline
\hline
$\Sigma_{c} \pi$ & 231.95 & 28.30\tabularnewline
$\Sigma_{c}^{*}\pi$ & 13.43 & 234.76\tabularnewline
Total width & 245.38 & 263.06\tabularnewline
\hline\hline
\end{tabular*}
\end{center}
\end{table}

\section{summary}{\label{summary}}
In this work, we study the strong decay behaviors of the newly observed resonance $\Lambda_b(6072)$ by the LHCb Collaboration. Given its mass and decay modes, this resonance can be tentatively assigned as the $\Lambda_b(2S)$, $\Sigma_b(1P)$ and $\tilde{\Lambda}_b(1P)$ states. The strong decay behaviors are investigated within the $^3P_0$ model, and our results suggest that the $\Lambda_b(6072)$ can be regarded as the lowest $\rho-$mode excitation in $\Lambda_b$ family.

Then, the strong decays for the nonstrange partners of $\Lambda_b(6072)$ resonance are calculated. The predicted total decay widths of the $J^P=5/2^-$ $\Lambda_b$ and $\Lambda_c$ states are relatively small, and the main decay channels are the $\Sigma_b^{(*)} \pi$ and $\Sigma_c^{(*)} \pi$, respectively. These two states have good potentials to be observed in future experiments, which may help us to distinguish the three-quark model and diquark model.

Until now, there exist abundant $\lambda-$mode states in the heavy baryon systems, while no $\rho-$mode excitaion has been confirmed both theoretically and experimentally. Under our assignment, the $\Lambda_b(6072)$ should correspond to the $\rho-$mode excitation for the singly heavy baryons. We hope these theoretical predictions on $\tilde{\Lambda}_{b(c)}(1P)$ and $\tilde{\Sigma}_{b(c)}(1P)$ states can provide valuable information for searching more $\rho-$mode excitations in future experiments.

\bigskip
\noindent
\begin{center}
{\bf ACKNOWLEDGEMENTS}\\

\end{center}
We would like to thank Xian-Hui Zhong for helpful suggestions. This project is supported by the National Natural Science Foundation of China under Grants No. 11705056 and No.~U1832173.

\end{document}